\documentclass[a4paper,fleqn]{cas-dc}

\usepackage{amsmath}
\usepackage{braket}
\usepackage[numbers]{natbib}
\usepackage{placeins}

\begin{document}

\title[mode=title]{Three ways to share a QPU: Scheduling strategies for hybrid Quantum-HPC applications}
\shorttitle{Three ways to share a QPU}
\shortauthors{M. Cipollini, S. Rizzo et al.}

\author[1,2]{Marco Cipollini}
\cormark[1]
\author[3]{Simone Rizzo}
\cormark[1]
\author[4]{Sergio Iserte}
\author[2]{Paolo Viviani}
\cormark[2]
\ead{paolo.viviani@linksfoundation.com}
\author[1,2]{Giacomo Vitali}
\author[3,8,9]{Matteo Barbieri}
\author[3]{Gabriella Bettonte}
\author[3]{Elisabetta Boella}
\author[3]{Fulvio Ganz}
\author[3]{Roberto Rocco}
\author[3,7]{Orazio Spina}
\author[4]{Antonio J. Peña}
\author[4]{Petter Sandås}
\author[5]{Iacopo Colonnelli}
\author[2]{Alberto Scionti}
\author[1,2]{Chiara Vercellino}
\author[2]{Emanuele Dri}
\author[6]{Jonathan Frassineti}
\author[6]{Sara Marzella}
\author[6]{Andrea Muratori}
\author[3]{Daniele Ottaviani}
\author[2]{Olivier Terzo}
\author[1]{Bartolomeo Montrucchio}
\author[3]{Daniele Gregori}

\affiliation[1]{organization={Politecnico di Torino},
                city={Torino},
                country={Italy}}
\affiliation[2]{organization={LINKS Foundation},
                city={Torino},
                country={Italy}}
\affiliation[3]{organization={E4 Computer Engineering SpA},
                city={Scandiano},
                country={Italy}}
\affiliation[4]{organization={Barcelona Supercomputing Center (BSC)},
                city={Barcelona},
                country={Spain}}
\affiliation[5]{organization={Università di Torino},
                city={Torino},
                country={Italy}}
\affiliation[6]{organization={CINECA},
                city={Casalecchio di Reno},
                country={Italy}}
\affiliation[7]{organization={Università di Bologna},
                city={Bologna},
                country={Italy}}
\affiliation[8]{organization={Dipartimento di Fisica e Astronomia "G.                          Galilei",  Universit\`a di Padova},
                city={Padova},
                country={Italy}}
\affiliation[9]{organization={Istituto Nazionale di Fisica Nucleare (INFN)},
                city={Padova},
                country={Italy}}

\cortext[cor1]{Equal Contribution}
\cortext[cor2]{Corresponding author}


\begin{abstract}
As quantum computing (QC) technologies mature, their integration into established high-performance computing (HPC) infrastructures is becoming a central objective for next-generation computing systems. However, unlocking the potential of hybrid platforms for computationally demanding workloads remains challenging. The mismatch between quantum and classical programming models, the limited maturity of quantum software stacks, and the scarcity of quantum processing units (QPUs) above all, necessitate scheduling strategies that go beyond standard HPC mechanisms to manage such heterogeneous and constrained resources.

To address this issue, we investigate three distinct methodologies for HPC-QC resource scheduling: time-based multiplexing, dynamic resource management, and workflow decomposition. 
Experimental validation on production HPC clusters and real quantum hardware demonstrates the effectiveness of these approaches under different workload scenarios. Malleability and workflow strategies significantly optimise classical resource utilisation, reducing consumption by up to 45.7\% and 64\% respectively, proving to be best fitted for hybrid jobs where quantum and classical workloads are evenly balanced. Conversely, time-multiplexing enhances QPU utilisation and reduces execution time at the cluster level, making it the optimal strategy for the opposite context, which is characterised by high classical-quantum workload imbalances. These findings underscore the practical viability of tailored scheduling strategies for hybrid HPC-QC environments and highlight their complementarity in building efficient, scalable software stacks for next-generation quantum-accelerated facilities.
\end{abstract}

\maketitle

\section{Introduction}\label{sec:intro}
Quantum computing (QC) is evolving rapidly, strongly establishing itself as a cornerstone in the future of advanced technologies. In this context, integration with traditional high-performance computing (HPC) systems is regarded as a crucial step towards fully exploiting its potential~\cite{EuroHPC:24,michielsenETP4HPCSRA62025}, leveraging the unique capabilities of QC to accelerate computationally intensive and exponentially scaling workloads while leaving the rest to the classical system. However, such integration raises new challenges on both sides, potentially hindering the development of both if left unchecked.

Developing an application that runs on a hybrid HPC-QC cluster is complex: the classical and quantum parts require radically different approaches to application design, and the gap in the maturity of the tooling used in the two fields, as well as the lack of standards, makes an effective integration not trivial. Additionally, there is a substantial imbalance between the availability of quantum and traditional resources, as quantum computers are still very scarce. This scarcity poses a significant challenge to the integration and resource allocation process of near-term quantum computers, which must account for it to prevent introducing bottlenecks and under-utilisation throughout the entire system. On the other hand, many of these systems are being integrated in HPC facilities at the time of writing~\cite{lrz_integration}, and the first extended software solutions have begun to emerge, providing unified interfaces to connect heterogeneous quantum hardware with classical infrastructures \cite{mqss}. Consequently, addressing the complexities of HPC-QC resource scheduling has become an issue of immediate practical concern.

The approaches presented in this paper revolve around a key observation: solutions to this challenge should not require HPC centres to overhaul their existing infrastructure or policies. HPC providers operate mature software stacks, including workload managers, schedulers, job accounting, and access control frameworks, that have been refined over decades. Any viable approach to integrate near-term, scarce quantum resources within HPC facilities must be deployable within these constraints, augmenting rather than replacing what is already in place.

In this context, this work discusses three methodologies: \emph{quantum resource multiplexing}, \emph{dynamic resource management}, and \emph{workflow decomposition}. Dynamic resource management and workflow decomposition operate at the level of the application's programming and execution model: the scheduler and the workflow management system (WMS) are responsible for adapting resource assignments at runtime, exploiting MPI malleability and independent allocations, respectively. Crucially, these mechanisms can be surfaced to the user through familiar HPC abstractions, such as malleable job submissions and workflow descriptions, without exposing the underlying quantum resource management at all, but treating quantum processing units (QPUs) as an HPC accelerator managed by established batch schedulers. Quantum resource multiplexing, on the other hand, operates at the infrastructure level: virtual QPUs (vQPUs) allow a single physical quantum processor to be shared transparently among multiple concurrent jobs by using well-known HPC scheduling tools. From the user's perspective, a vQPU is a quantum resource to target in their batch script, while time slicing is handled transparently behind the scenes. Potentially, a user can leverage workflow decomposition or job malleability without being aware that the targeted QPU is also being shared with other jobs.

Any proposed approach can be classified and analysed along different orthogonal axes. This article focuses on two: the \emph{time scale} (i.e., the granularity of quantum workloads and resource allocation) and the \emph{programming model} (i.e., how much the application must be aware of and explicitly structured around hybrid resource allocation). We present these integration strategies together, as they fit different regions of these two axes, complementing each other and potentially covering a broad class of workloads and operational scenarios.

Building on previous work~\cite{Simone,QCE_malleability_2025}, this article provides experimental validation to the quantum resource multiplexing mechanism, reporting results from a test campaign on a production HPC cluster and a real quantum machine. It also extends the experimental results on workflow decomposition and dynamic resource management to a production environment. These new results further underscore the practical relevance of the proposed approaches and close the gap between theoretical analysis and experimental validation on real infrastructure. In summary, the contributions of this work are the following:

\begin{itemize}
    \item We consolidate three strategies for efficient HPC-QC resources scheduling, clarifying and extending their conceptual framework.
    \item We present the design and results of an experimental campaign aimed at evaluating three scheduling strategies across quantum jobs with different time scales. These experiments realistically replicate real-world conditions, where quantum machines operate with significantly diverse temporal characteristics. Finally, we extend the validation of our methodology to production HPC clusters and real quantum hardware, further reinforcing the practical viability of all three proposed approaches.
    \item Finally, we discuss the strengths, weaknesses, and the domain of applicability of each strategy, providing both users and HPC-QC operators with best practices and guidelines to deploy these strategies in their specific scenario.
\end{itemize}

This work is structured as follows: Section~\ref{sec:sota} reports on the current state of the art of Quantum and HPC integration; Section~\ref{sec:methodology} discusses the conceptual framework behind the three proposed strategies; Section~\ref{sec:exp} presents the test cases and the experimental set-up used to demonstrate the effectiveness of the three strategies; finally, Section~\ref{sec:summary} provides an overview of the results, discussing the complementarity of the three proposed methodologies, their applicability and the future perspectives.

\section{Background and related work}\label{sec:sota}
HPC-QC integration is an active research topic~\cite{schulzAcceleratingHPCQuantum2022a,humbleQuantumComputersHighPerformance2021,bartschvaleriaQCHPCQuantum2021,brittHighPerformanceComputingQuantum2017,mccaskeyXACCSystemlevelSoftware2020, garciaraigada2026wavebaseddispatchcircuitcutting}, promoted by hardware vendors~\cite{seelam2026referencearchitecturequantumcentricsupercomputer,ruefenachtBringingQuantumAcceleration} and supported by institutions such as the European Commission through the EuroHPC Joint Undertaking.\footnote{\url{https://doi.org/10.3030/101018180}} These efforts have significantly contributed to the classification of quantum applications~\cite{humbleQuantumComputersHighPerformance2021}, the definition of HPC-QC architectures~\cite{bartschvaleriaQCHPCQuantum2021,humbleQuantumComputersHighPerformance2021,beck2024quantum}, the definition of coherent software stacks~\cite{mqss,humbleQuantumComputersHighPerformance2021,schulzAcceleratingHPCQuantum2022a,willeMQT}, and the initial outline of programming models~\cite{mccaskeyXACCSystemlevelSoftware2020, schulzAcceleratingHPCQuantum2022a}.

However, although quantum machines are increasingly being installed close to HPC facilities, they are still primarily accessed through high-level languages and dedicated libraries (e.g., Qiskit, Pulser). These tools allow users to define kernels (i.e., circuits) and submit them via REST APIs, typically resulting in sequential, mostly single-threaded executions. While this cloud-like access model is effective for using current quantum machines, it falls short of true HPC integration, which demands direct, high-bandwidth connectivity and closed-loop control over the quantum hardware.

In order to provide the right context for a discussion, we need a taxonomy of HPC-QC integration modalities. Building on previous literature~\cite{Esposito_2023,Elsharkawy_2024,shehata_2025,mantha_2024,beck2024quantum,humbleQuantumComputersHighPerformance2021,bartschvaleriaQCHPCQuantum2021,Saurabh_middleware_2023}, our previous work~\cite{Simone} identified three main modalities:
\begin{itemize}
    \item \emph{Tightly coupled}: classical computation happens within the coherence time of the quantum machine, and affects the execution of the circuit. The main use case for this is error correction (i.e., by mid-circuit measurements, syndrome extraction, and decoding), which currently represents more a quantum HW challenge than a matter of HPC-QC integration.
    \item \emph{Quantum offloading}: typical circuit execution is used to accelerate specific operations (i.e., quantum kernels). It can be either synchronous or asynchronous, but the classical code manages the control flow and continues running until completion.
    \item \emph{Loosely coupled}: an external controller (as simple as a shell script or as complex as an entire workflow manager) handles the control flow. Classical and quantum resources are allocated separately for each step once that step is ready to run.
\end{itemize}
The first mode still requires advancements in HW maturity and it pertains primarily to error correction. Therefore, it is outside the scope of this work, which focuses instead on practical, near-term challenges for the second and third integration modes. The same work~\cite{Simone} also introduced first the three-pronged strategy discussed and extended here, encompassing QPU multiplexing, dynamic resource management and workflow decomposition. The latter two strategies have been also analysed in a later work~\cite{QCE_malleability_2025} through an experimental campaign on a small cluster featuring an example hybrid application, demonstrating the significant benefits achievable through intelligent resource assignment. In the following sub-sections we will discuss the state of the art of specific aspects of HPC-QC integration that are relevant in this context.

\subsection{Applications}
The synergy between HPC and QC holds significant potential to accelerate progress across multiple scientific domains. Building on insights from foundational studies~\cite{beck2024quantum, adac,seelam2026referencearchitecturequantumcentricsupercomputer}, hybrid HPC–QC systems are expected to have the greatest impact in areas such as condensed matter physics~\cite{material_science}, quantum chemistry~\cite{chemistry} and, in the long-term, combinatorial optimisation and artificial intelligence (AI)~\cite{quantum_ml}. The literature already reports concrete implementations of hybrid algorithms evaluated both on large-scale HPC simulators and on early HPC–QC prototype platforms.
For example, Kim~\textit{et~al.} improved the execution of the Quantum Approximate Optimization Algorithm (QAOA) for materials science applications using the MPI programming model~\cite{kim_2024}, and further experimented with real quantum hardware loosely coupled to a distributed classical system~\cite{DQAOA}. Similarly, Vercellino~\textit{et~al.} integrated a quantum subroutine into a parallel branch-and-bound framework to address the graph coloring problem \cite{Vercellino_2023}, with the potential to execute nodes in the same layer of the search tree as parallel MPI processes, each one calling a quantum subroutine. Beyond hybrid algorithms, HPC infrastructures also play a crucial role in supporting large-scale quantum circuit simulations. In this context, Shang~\textit{et~al.} developed a simulator based on Matrix Product State (MPS)~\cite{vqe_hpc}, capable of efficiently executing the Variational Quantum Eigensolver (VQE) algorithm, leveraging parallel classical resources to tackle challenging quantum chemistry problems. Complementing this line of work, Hamamura~\textit{et~al.} recently proposed an implementation of the Quantum-Selected Configuration Interaction (QSCI) class of quantum-classical methods, combining GPU-accelerated quantum simulations with large-scale CPU-based subspace diagonalisation \cite{nvidia_hybrid}.

\subsection{QPU sharing}\label{sec:sota-vqpu}
The promising results discussed in the previous section reinforce the appeal of heterogeneous HPC–QC clusters. However, their practical deployment remains constrained by several structural challenges, including the different levels of technological maturity of HPC and QC systems, their fundamentally distinct architectural paradigms, and the pronounced asymmetry between classical and quantum resource availability \cite{Elsharkawy_2023,viviani_keynote}. In this scenario, efficient resource management and allocation become central concerns \cite{Elsharkawy_2023, mantha_2024, shehata_2025, dobler2025surveyintegratingquantumcomputers}. Effective scheduling policies are essential to prevent resource under-utilisation, particularly in configurations where only a limited number of QPUs are available within the cluster \cite{schulz_2022,brittHighPerformanceComputingQuantum2017,saurabh_2023}.
Previous literature~\cite{zhan2025stackframeworkhighperformance,esposito2025Slurmheterogeneousjobshybrid} proposed splitting Slurm jobs into building blocks to allow overlapping classical and quantum tasks among different jobs, to allow better QPU utilisation, while other authors proposed QPU partitioning~\cite{multi-programming,qu_cloud}, or a full fledged virtualisation approach, abstracting the hardware to enable qubit-based QPU sharing~\cite{ZHENG2026_vqpu}. Further approaches to QPU multi-tenancy~\cite{liu2026dynqdynamictopologyagnosticquantum} also consider the qubit topology and noise structure to identify optimal regions for mapping multiple quantum jobs on the same QPU. Marginally related to quantum resource scheduling, we mention the CUNQA library~\cite{vazquezperez2025cunqadistributedquantumcomputing}, that also introduces the virtual QPU term, albeit with a different definition.

\subsection{Dynamic resource management}
Dynamic resource management denotes mechanisms that allow a running job to change the amount and type of resources it uses while executing, instead of maintaining a fixed allocation decided at submission time.
In HPC systems, it enables jobs to adapt to heterogeneous nodes or changing performance conditions without stopping and restarting the application, for instance, by expanding jobs when spare nodes are available, or shrinking them under load or energy constraints.

Within this broader dynamic resource management space, MPI malleability focuses on applications that use MPI and can adjust their process layout at runtime~\cite{feitelson_packing_1996}. A malleable MPI application can add or remove MPI processes during execution, in coordination with the resource management system and the MPI runtime, while preserving correct communication and data distribution. This requires tightly coupled interactions between the application, a malleability library, the MPI process manager, and the batch scheduler.

MPI malleability has been widely studied in the literature~\cite{aliaga_survey_2022, tarraf_malleability_2024}.
Several solutions have explored the effects of dynamic jobs within HPC workloads from benchmarks~\cite{sudarsan_reshape_2007, prabhakaran_batch_2015}, synthetic setups~\cite{iserte_dmrlib_2020}, or simulations~\cite{sarood_maximizing_2014} to actual HPC log replaying~\cite{iserte_mpi_2025}, showcasing the benefits regarding waiting times, makespan, and resource utilisation~\cite{iserte_agut_high-throughput_2018}.
Although traditionally malleability tools have been evaluated with ad-hoc schedulers~\cite{sudarsan_dynamic_2009, huber_bridging_2025} or customised resource managers~\cite{Compres2016, iserte_resource_2025}, recent frameworks such as DMR have jumped in to the production environments, allowing dynamic jobs to run in standard partitions and enabling malleability without interfering with traditional rigid jobs~\cite{iserte_malleable_2025}.

\subsection{Workflow decomposition}
Another approach to the HPC-QC scheduling challenge involves the use of workflow systems~\cite{osti_2006942}, which provide higher-level orchestration mechanisms for heterogeneous workloads. Empirical demonstrations of hybrid HPC–QC applications effectively leveraging WMSs to optimise resource distribution have already been reported \cite{CRANGANORE2024346, DeMaio_2024}. Our previous work~\cite{QCE_malleability_2025} explored the adoption of StreamFlow \cite{Colonnelli:2021} to enable seamless task coordination, thereby promoting workflow-centric architectures in the HPC–QC domain. More recently, a Python-based workflow framework has been introduced to manage hybrid applications across HPC systems and quantum computers connected via a shared network by integrating the PSI/J library with custom job submission scripts~\cite{python_workflow} In another recent work, Tejedor~\textit{et~al.} \cite{kubernetes_hpc_qc} introduced a cloud-native framework for orchestrating hybrid quantum-classical workflows by integrating Kubernetes and Kueue software platforms with the Argo Workflows WMS. The system provides a unified orchestration layer that manages heterogeneous resources, including CPUs, GPUs, and QPUs, enabling multi-stage pipelines with dynamic, resource-aware scheduling.


\section{Resource allocation strategies}\label{sec:methodology}

\begin{figure}
    \centering
    \includegraphics[width=\linewidth]{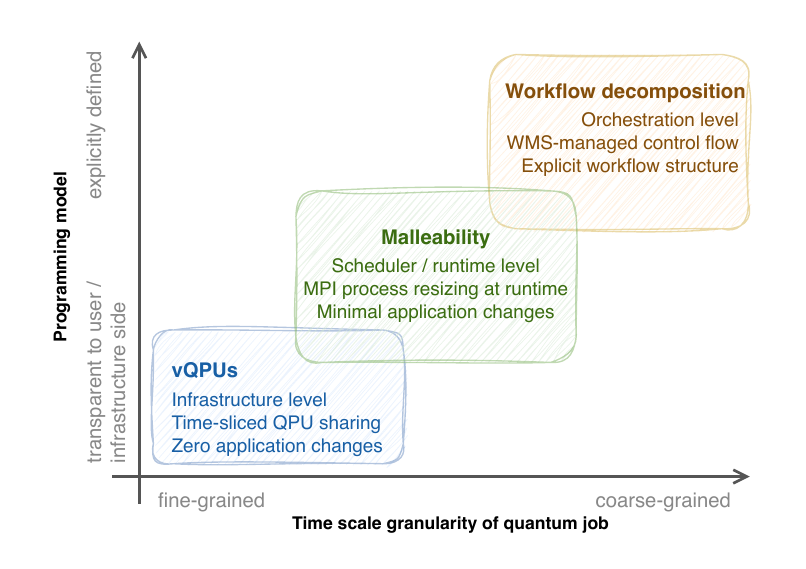}
    \caption{Positioning of the three proposed scheduling strategies along two orthogonal axes: the time-scale granularity at which quantum and classical workloads alternate (horizontal) and the degree of programming-model transparency, i.e., the extent to which the application must be explicitly restructured to benefit from the strategy (vertical).}
    \label{fig:overview}
\end{figure}
This section presents the three complementary resource allocation strategies depicted in Figure~\ref{fig:overview}, each targeting a different scheduling granularity and a different layer of the software stack. These strategies apply to NISQ devices, but can also affect future error-corrected devices, where the classical compute needed for error correction is packaged together with the quantum hardware. Moreover, addressing different time scales and granularities, allows to handle efficiently different qubit modalities with different time scales~\cite{Simone}.
At the infrastructure level, quantum resource multiplexing through virtual QPUs enables multiple concurrent jobs to share a single physical QPU transparently, with no changes required to user applications. At the cluster level, dynamic resource management leverages MPI malleability to adapt classical process allocations during execution, releasing idle nodes while quantum phases run. At the application level, a workflow models a complex problem as a graph of tasks, delegating resource allocation, orchestration, and fault tolerance to a WMS. These three strategies are best understood along the two orthogonal axes introduced in Section~\ref{sec:intro}, i.e., the time scale and the programming model. As illustrated in Figure~\ref{fig:overview}, they occupy distinct but partially overlapping regions of this space, making them applicable to different classes of workloads and operational scenarios — and, in some cases, complementary within the same set-up.
This potential composability among the proposed approaches was not explored in our experimental campaign and is deferred to future work, as discussed in Section~\ref{sec:future}.

\subsection{VQPUs}\label{sec:methods_vqpu}
Given the current characteristics of the available superconducting hardware and of typical quantum-accelerated applications, where quantum computations typically represent short execution bursts interleaved with longer classical processing phases, it is particularly relevant to consider scenarios in which the quantum contribution occupies only a marginal portion of the total runtime, while other qubit technologies and scenarios deserve a separate discussion.

\begin{figure*}
    \centering
    \includegraphics[width=0.8\linewidth]{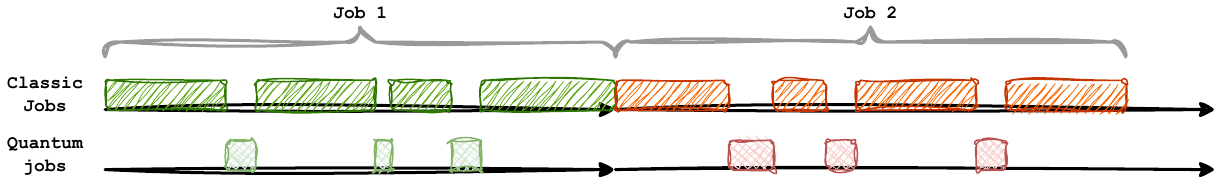}
    \caption{An illustrative example of the execution of two hybrid jobs, following the naïve co-scheduling policy. The upper/lower row represents the timeline of classical/quantum computations. Since there is only one available QPU in the system, the first submitted job holds it for its entire duration, preventing the second one to even begin its classical computations. This ultimately leads to a major under-utilisation of the quantum resource overall.}
    \label{fig:coscheduling}
\end{figure*}

Such a situation is ideally represented in Figure \ref{fig:coscheduling}, where jobs are submitted under a naïve co-scheduling model, from now on referred to just as \textit{co-scheduling}, to a computing system with only one available QPU. In this setting, each user obtains exclusive access to the QPU during the duration of their job. Due to the pronounced workload imbalance between classical and quantum computations, the QPU remains idle for most of the execution time, leading to suboptimal utilisation of the scarcest resource in the system. This inefficiency emphasises the importance of treating QPUs as a shared infrastructure within HPC environments and motivates the development of mechanisms that allow for concurrent access among multiple hybrid applications.

A possible solution to mitigate QPU contention could involve partitioning and assigning subsets of qubits to different concurrent jobs. However, this approach requires to carefully compile multiple jobs together into a single pulse-level job, taking into account potential crosstalk and connectivity issues, possibly along with additional gates; in our opinion, while interesting, an operationalisation of such approach remains an open research topic.

A more practical and promising strategy relies on time-based multiplexing, which enables multiple applications to share the same physical QPU by alternating their quantum executions. In this model, the QPU is shared over time, allowing each application to perform its quantum operations within distinct temporal windows. This concept can be effectively realised through QPU virtualisation~\cite{Simone}, where a fixed or dynamic number of virtual QPUs (vQPUs) are defined, each corresponding to a fraction of the QPU's total runtime rather than to a physical partition of the device.

\begin{figure}
    \centering
    \includegraphics[width=0.8\linewidth]{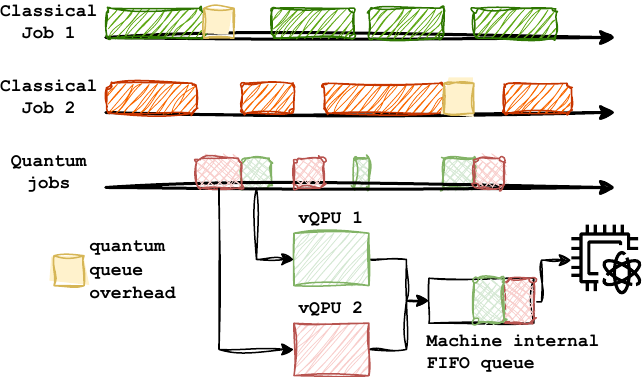}
    \caption{A representation of the execution of the same two hybrid jobs in Figure \ref{fig:coscheduling}, now following the time-based multiplexing scheduling policy. Since they are both granted a time-share of the QPU, the two jobs are allowed to start their classical computations at the same time, assuming they were submitted simultaneously. The QPU will then process the incoming quantum phases from the two jobs sequentially, according to its internal queue. If the imbalance between quantum and classical workloads' duration is substantial, this policy causes only a small quantum overhead to each job's duration, while the overall execution time at cluster level is significantly reduced.}
    \label{fig:vqpu_scheduling}
\end{figure}

Through this mechanism, applications can request one vQPU, possibly managed by well-established job scheduler directives like Slurm GRES (\verb|--gres=qpu:1|) or Slurm licenses (\verb|--licenses=qpu:1|), while, in reality, their executions are interleaved on the same physical hardware. As illustrated conceptually in Figure \ref{fig:vqpu_scheduling}, multiple hybrid jobs can concurrently issue quantum execution requests, which are then scheduled asynchronously and run sequentially on the QPU according to its internal queue. The main advantage of this strategy lies in the fact that it increases overall QPU availability without requiring any modification to the user applications, which remain unaware of the multiplexing layer.

Nevertheless, the effectiveness of this approach is strongly influenced by the ratio between classical and quantum workloads' duration. When the quantum portion of a job becomes comparable to or exceeds the duration of the classical part, the benefits of virtualisation decrease, as interleaving can no longer effectively mask QPU idle times. Therefore, the vQPU model is most advantageous in hybrid workloads characterised by a significant classical-to-quantum imbalance, where short quantum computations can be efficiently overlapped with classical processing, maximising resource utilisation at the cluster level. This scenario is typical of superconducting quantum machines integrated in HPC environments.

It is important to highlight that the adoption of vQPUs inherently involves accepting and balancing two distinct types of trade-offs. The first concerns the quantum–classical optimisation trade-off: as illustrated by the comparison between Figures \ref{fig:coscheduling} and \ref{fig:vqpu_scheduling}, employing vQPUs entails sacrificing a small fraction of optimisation in the abundant classical resources to achieve a substantial improvement in the utilisation of the scarce quantum ones. In other words, the approach prioritises the efficient exploitation of the QPU, the cluster's most limited, even at the cost of a minor inefficiency in the classical domain.

The second trade-off relates to the balance between cluster-level optimisation and individual user experience. By enabling shared access to the QPU, the vQPU model improves global throughput and overall resource utilisation within the cluster. However, since quantum resources are concurrently accessed by multiple jobs and QPU preemption is not currently feasible, individual users may occasionally experience longer execution times for their own job, potentially degrading their perceived performance.
When these aspects are considered from a cluster-oriented perspective, the advantage of adopting vQPUs becomes evident: the overall efficiency of quantum resource allocation is significantly enhanced, ultimately benefiting the HPC-QC system as a whole.

\subsection{Malleability}\label{sec:methods_malleability}


While cloud execution environments already feature malleability properties, the HPC scenario still relies on static allocation. This limitation comes mainly from two reasons: first, the focus towards individual application performance more than cluster-wide execution efficiency, as resource reallocation may cause overheads in the execution; second, MPI (the de-facto standard for inter-process communication) lacks proper dynamic process management, thus HPC applications tend to not use it. However, in an HPC-QC integration scenario, the scarcity of quantum resources, as well as the difference in typical time scales, calls for a wider adoption of malleable jobs through dedicated libraries and HPC facility-level support, to provide better cluster-wide occupancy and shorter queue times for all users.

Among the several tools available to aid the implementation of malleability~\cite{aliaga_survey_2022}, the DMR framework~\cite{iserte_dmrlib_2020} provides a high-level API designed to incorporate malleability into HPC applications seamlessly.
DMR is a well-known tool in the community~\cite{tarraf_malleability_2024}, thoroughly evaluated across a variety of use cases~\cite{iserte_study_2020, iserte_dynamic_2018, iserte_dmr_2018}, and has also been extended with novel functionalities to support broader capabilities~\cite{huber_bridging_2025, iserte_resource_2025}.

At its core, DMR enables passive dynamic resource management through the following process:
\begin{enumerate}
\item During execution, DMR periodically checks when jobs are ready for reconfiguration at predefined synchronisation points within the application.
\item DMR determines whether a reconfiguration is necessary, considering available resources and performance metrics.
\item DMR orchestrates the updates in resource allocations (interacting with the resource management system) and the number of processes (through the parallel distributed runtime).
\item Finally, the execution resumes seamlessly from the reconfiguration point with the new job shape.
\end{enumerate}

Through the use of DMR, hybrid HPC-QC applications can release the resources they are not using, ensuring better utilisation for the overall system without compromising the results of each singular job. Our previous work~\cite{QCE_malleability_2025} analysed the use of DMR in a limited experimental scenario, while this article extends the evaluation to a production cluster.

\subsection{Workflows}\label{sec:methods_wf}
A workflow approach decomposes an application into a set of tasks and executes them according to the dependencies between them~\cite{Suter:2026}. The \emph{hybrid workflow} paradigm generalises this model by enabling the steps of a single workflow to span multiple, heterogeneous, and independent computing environments \cite{Colonnelli:2023}. Formally, a hybrid workflow can be represented as a triple $(W,$$L,$$\mathcal{M})$, where $W$$=$$(S,$$P,$$\mathcal{D})$ is a directed bipartite graph that captures the structure of a standard workflow (with steps $S$, ports $P$, and dependency links $\mathcal{D}$), $L$ is the set of available execution locations, and $\mathcal{M}$$\subset$$(S$$\times$$L)$ is a mapping relation that associates workflow steps with execution locations \cite{Colonnelli:2024}.

This abstraction is general enough to encode workflow models targeting quantum-suitable applications, dramatically facilitating their evolution from purely classical execution to HPC-QC settings. For example, Cranganore \emph{et al.}~\cite{CRANGANORE2024346} require workflow designers to identify a set $Q$$\subseteq$$S$ of \emph{quantum candidates}, i.e., steps that support a quantum implementation that is \emph{functionally equivalent} to its classical counterpart. At runtime, the WMS selects between classical and quantum implementations, depending on the availability of quantum resources. This technique can be captured within a hybrid workflow by identifying a subset of \emph{quantum locations} $L_{Q}$$\subseteq$$L$ and defining a \emph{quantum mapping} function $\mathcal{M}$$\subseteq$$(Q$$\times$$L_Q)$. The principal advantage of the more general hybrid workflow abstraction is that it enables a seamless transition from specialised HPC-QC orchestrators to X-QC settings, including cloud-QC and HPC-cloud-QC.

Focusing on the specific HPC-QC scenario, delegating resource allocation to a WMS offers a key advantage: each execution location, whether classical or quantum, is provisioned only when a task is actively running. This allocation policy has two practical benefits. First, even when waiting times for quantum resources are long, classical resources are not kept idle. This avoids wasting compute time and reduces costs, thereby improving overall cluster efficiency, albeit at the cost of increased queue times. Second, if the majority of cluster users adopt workflow models for hybrid applications, a fair allocation of quantum resources is ensured by the cluster scheduler's policy. Other than that, workflows allows users to offload to the WMS control plane non-functional requirements like fault tolerance~\cite{Mulone:2026} and provenance collection~\cite{24:pone:wfrunrocrate}.

The StreamFlow WMS~\cite{Colonnelli:2021} has been explicitly designed to support the hybrid workflow paradigm. It augments the Common Workflow Language (CWL)~\cite{CWL:2022} open standard with a topology of deployment locations and a \texttt{Loop} extension to model iterative workflows~\cite{ml4astro:22}. The orchestration layer leverages pluggable connectors to support several execution environments, from HPC queue managers to cloud infrastructures. Developing a new plugin to support a given quantum device is just a matter of extending the StreamFlow \texttt{Connector} base class to integrate the device's APIs, whether via web-based REST APIs or Slurm-managed queues.

\section{Experimental validation}\label{sec:exp}
As discussed in Section~\ref{sec:methodology}, the three proposed strategies operate at largely orthogonal levels of the system stack: vQPUs address the scheduling of quantum resources at the infrastructure level, malleability optimises the dynamic scaling of cluster resources at the MPI program level, and workflow decomposition manages the orchestration of classical and quantum allocations at the whole application level (see Figure~\ref{fig:overview}). Since these strategies target different resources and mechanisms, they are complementary rather than competing: as such, they were largely developed around different experimental setups, each tailored to exercise the level of the stack that the corresponding strategy controls.

The first setup evaluates the vQPU time-multiplexing strategy by running multiple concurrent instances of a hybrid quantum-classical Graph Coloring solver, based on the BBQ-MIS algorithm~\cite{Vercellino_2023}, against a real superconducting QPU. The key experimental variable is the number of concurrent jobs contending for the physical QPU, and the metrics of interest, namely quantum occupancy, total execution time, and queue time, directly characterise the effectiveness of the multiplexing mechanism. A tunable artificial delay in the classical phase allows us to study different quantum-to-classical workload ratios, exploring different scenarios.

The second setup evaluates the malleability and workflow strategies by running a clustering-aggregation application~\cite{QCE_malleability_2025} whose quantum subroutine is emulated classically via Simulated Annealing. The focus here is on the efficient use of resources from the job perspective: the application's highly parallel classical component (three concurrent MPI-based clustering algorithms) dominates the runtime, while the quantum phase is short. The metrics of interest, wall time and node-seconds of classical resource usage, directly capture the gains that malleability and workflow orchestration provide over a static allocation baseline. Also in this case, we introduce artificial delay, this time in the quantum phase, to study different quantum-to-classical workload ratios mimicking both superconducting and neutral-atom scenarios. Emulating the quantum subroutine is not a limitation in this context, as these strategies are agnostic to how the quantum computation is performed; what matters is the temporal structure of the hybrid workload.

\subsection{Experimental Design and test cases}

\subsubsection{Graph Coloring}\label{sec:exp_vqpu}
To evaluate the effectiveness of the vQPU approach, it is essential to assess the cluster's performance under varying levels of concurrent hybrid workloads. Specifically, the experiment was designed to reproduce a worst-case scenario, in which all applications are of the same type, receive identical input data, and are launched simultaneously. This setup maximises the overlap in the submission of quantum circuits to the QPU and ensures that each job exhibits a similar quantum-to-classical workload ratio.

The selected benchmark problem for each application is the Graph Coloring (GC) problem \cite{graph_coloring}, which consists of finding the minimum number of labels (or colors) required to assign to the vertices of an undirected graph $G=(V, E)$ such that no two adjacent vertices share the same label. This combinatorial problem has several practical applications, including task scheduling and register allocation \cite{register_allocation}, and is well-known to belong to the NP-Hard class of computational problems. For this reason, exploring the use of quantum-accelerated approaches to address such computationally intensive problems is particularly well-motivated.

The experimental workflow is organised around a parameter space exploration over the number of GC instances that is submitted simultaneously using Slurm as job scheduler, emulating concurrent access to the QPU by multiple users. The experimental design assumes that the number of available vQPUs always matches the number of concurrent GC instances. This assumption is motivated by methodological considerations: if fewer vQPUs than submitted jobs were available, the scheduler would keep the excess jobs on hold until both quantum and classical resources were available, as it normally does. Conversely, allocating more vQPUs than the number of concurrent jobs would not alter the execution in any meaningful way. Ultimately, both cases do not provide additional insight into the experiment.

Each GC instance is solved using the hybrid branch-and-bound BBQ-MIS algorithm, originally introduced by Vercellino \emph{et al.}~\cite{Vercellino_2023}. Owing to the algorithm's inherent structural parallelism, multiple MPI workers can be allocated to accelerate the search for the optimal solution, making this method suitable to be executed in an HPC environment. However, the results presented in this article refer to runs with a single MPI worker, since executions with multiple workers did not lead to significant differences in the metrics under investigation.

The search for the optimal solution proceeds by decomposing the original problem into several Maximal Independent Set (MIS) subproblems~\cite{Vercellino_2023}, which can be, in turn, expressed as Quadratic Unconstrained Binary Optimization (QUBO) problems. Such subproblems can be addressed either through analogue quantum algorithms, e.g., Quantum Annealing, or through digital approaches. Given that a gate-based quantum computer was available for this study, each MIS problem was solved using the well-known Quantum Approximate Optimization Algorithm (QAOA)~\cite{qaoa}. In this framework, the combinatorial problem is encoded in a cost Hamiltonian~$H_C$, diagonal in the computational basis, whose ground state corresponds to the optimal solution. For MIS, $H_C$ takes the form of an Ising-type Hamiltonian that penalises violations of the independence constraint while rewarding large independent sets. The QAOA ansatz of depth~$p$ is then defined by the parametrised unitary $$U(\boldsymbol{\gamma},\boldsymbol{\beta})   = \prod_{k=1}^{p} e^{-i \beta_k H_M}\, e^{-i \gamma_k H_C},$$ where $H_M = \sum_i X_i$ is the transverse-field mixer Hamiltonian. A classical optimiser iteratively adjusts the $2p$ parameters~$(\boldsymbol{\gamma},\boldsymbol{\beta})$ so as to minimise the expectation value
$$
\langle H_C \rangle
  = \bra{+}^{\otimes n}\,
    U(\boldsymbol{\gamma},\boldsymbol{\beta})^\dagger\,
    H_C\,
    U(\boldsymbol{\gamma},\boldsymbol{\beta})\,
    \ket{+}^{\otimes n}
$$
where $\ket{+}^{\otimes n}$ is the uniform superposition obtained by applying~$H_M$ to the $n$-qubit initial state~$\ket{0}^{\otimes n}$. After convergence, a final measurement sampling is performed to extract the solution state of the corresponding MIS instance.

The variational circuit follows a modular structure: each alternated application of the cost and the mixing Hamiltonian is commonly referred to as a layer. The number of layers $p$ thus becomes a hyperparameter that must be balanced between approximation accuracy and parameter space dimensionality. In the experimental setup, the circuit was configured with a single layer. This choice was motivated by the limited relevance of solution quality to the objectives of the study, as well as by the application of the circuit cutting technique, to be discussed in the next paragraph, that would have dramatically increased the runtime of the jobs in a $p>1$ scenario.

The variational circuits generated by QAOA were submitted to the quantum computer via a dedicated API, which handled all incoming requests sequentially across the different GC instances running in parallel thanks to its internal First In First Out (FIFO) queue. In addition to submitting quantum circuits to the QPU, the BBQ-MIS framework involves several post-processing steps and the application of pruning strategies that are executed exclusively on the classical nodes of the cluster, thereby making the method genuinely hybrid in nature. To investigate the metrics of interest in this study as a function of the ratio between classical and quantum workload, a sleep mechanism was incorporated into the algorithm in order to artificially increase the classical execution time.

\paragraph{Circuit Cutting.}
Given the limited number of qubits available on the quantum device, the \textit{circuit cutting} technique was employed to enable the execution of circuits involving more than 5 qubits \cite{cutting_1}. Circuit cutting allows a quantum circuit to be decomposed into smaller sub-circuits, each with fewer qubits, that can be executed independently one from the other. The results are later recombined through classical post-processing to achieve an approximation of the original circuit's outcome. In particular, this method is designed to compute the expectation value of a generic observable $\braket{O}$ over the original circuit, which aligns naturally with the QAOA workflow.

Two main cutting strategies are typically considered: wire cutting, which partitions the circuit by severing qubit lines and inserting \textit{measure-and-prepare} operations at the cut edges; and gate cutting, which decomposes two-qubit gates into probabilistic mixtures of single-qubit operations \cite{cutting_2}. It is important to note that both strategies introduce an exponential overhead in the number of sub-circuits that must be executed with respect to the number of cuts applied, which limits their applicability to specific circuit topologies where cuts can be performed efficiently.

\begin{figure}
    \centering
    \includegraphics[width=0.9\linewidth]{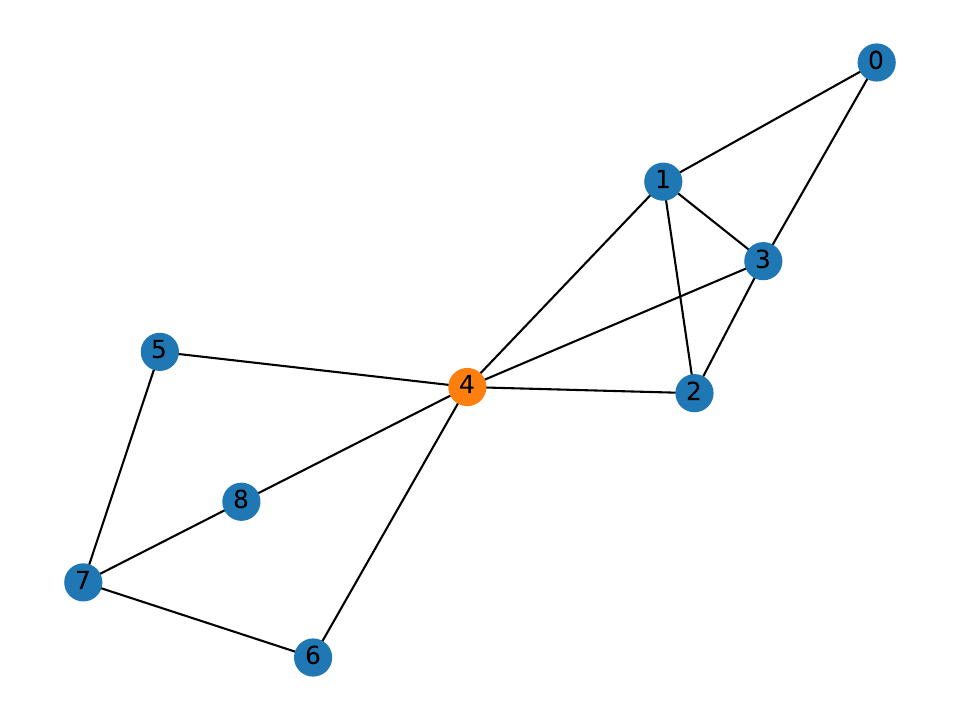}
    \caption{A (1,5)-clustered graph used as input to the GC problem. The central node (label 4) is the vertex separator of 2 clusters with 5 nodes each.}\label{fig:clustered_graph}
\end{figure}

Previous studies have explored the application of circuit cutting within the QAOA framework \cite{qaoa_cutting_1, qaoa_cutting_2}. These works focused on combinatorial graph problems, such as Max-Cut, and in particular on a class of graphs known as \textit{(K,d)-clustered graphs}. A graph $G=(V,E)$ is defined as \textit{(K,d)-clustered} if there exists a subset of vertices $S$, referred to as \textit{vertex separators}, such that $|S|=K$ and the reduced graph $G' = G \setminus S$ consists only of connected components containing at most $d$ vertices. Under this definition, the decomposition of $G$ into smaller connected components requires only a minimal number of cuts, each corresponding to a sub-circuit involving at most $d$ qubits. The present experiment adopts the same class of graphs, with $d=5$, an example of which is provided in Figure \ref{fig:clustered_graph}.
Since its original formulation, several implementations of circuit cutting have been proposed, each characterized by specific requirements and potential advantages in terms of overhead reduction \cite{cutting_2}. The most effective approaches typically rely on establishing \textit{classical communication} among sub-circuits, allowing information to be exchanged during runtime. As this requirement could not be fulfilled in our setup, the original wire cutting method described in \cite{cutting_1} was adopted, with no classical communication involved. To further limit the number of cuts applied to the circuit, QAOA was configured with a single layer, as outlined in Section \ref{sec:methods_vqpu}.

Moreover, since the workflow required not only the evaluation of the expectation value $\braket{H_C}$ during parameter training but also a final sampling of the circuit to extract the MIS solution, an operation not originally addressed in the circuit cutting formulation, the sampling strategy proposed in \cite{qaoa_cutting_2} was employed. All circuit cutting procedures were implemented using the Python library \verb|qiskit_addon_cutting|, with the exception of the sampling method, which was developed from scratch.

\subsubsection{Clustering aggregation}
Building on the work of Scotti \textit{et al.}~\cite{scotti2024clusteringaggregationalgorithmneutralatoms} and Li \textit{et al.}~\cite{Li2012ClusteringAA}, we introduced a new hybrid HPC–QC application, which we use to assess the effectiveness of the workflow decomposition and malleability strategies~\cite{QCE_malleability_2025}. The algorithm performs clustering aggregation to determine the optimal number of clusters based on the outputs of multiple clustering methods.
This application was specifically designed to support our study of dynamic quantum–HPC resource management: its classical component is highly parallelisable and can efficiently leverage concurrent execution, while its need for quantum resources is restricted to a short portion of the overall runtime.
In this work, we leverage the HPC–QC clustering‑aggregation algorithm to investigate the potential of both malleability‑based approaches and workflow‑oriented strategies for efficient resource allocation, evaluating them on both a testbed and a production‑scale cluster.
\begin{figure}
    \centering
    \includegraphics[width=0.7\columnwidth]{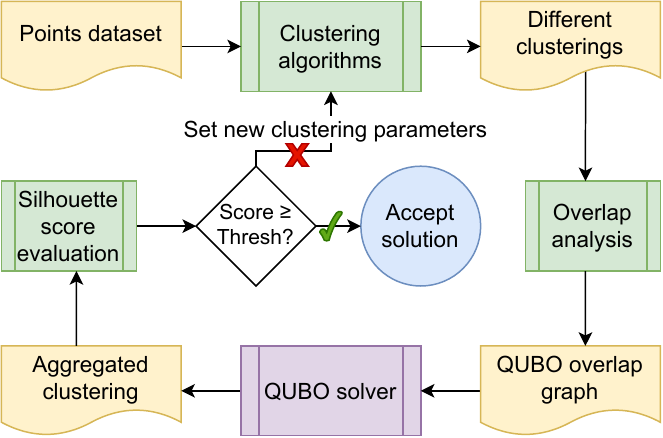}
    \caption{Data and processes of the target application. Yellow shapes are data between processes, green boxes are classical processes, and purple boxes are processes suitable to be performed by quantum resources. The application loops until the Silhouette score of a certain clustering reaches a user-defined threshold or until the maximum number of loop iterations is reached.}
    \label{fig:proposed_algorithm}
\end{figure}

Our previous work~\cite{clustering-mis, QCE_malleability_2025} introduced a C++ application that executes three widely used clustering algorithms (i.e., k-means~\cite{kmeans}, DBSCAN~\cite{dbscan}, and hierarchical clustering~\cite{hclust}) in parallel on top of separate HPC nodes using MPI (see Figure~\ref{fig:proposed_algorithm}). This parallelisation yields significant performance gains over the serial baseline when multiple compute nodes are available. The outputs of the three algorithms are then combined to construct an undirected graph under the constraint that clusters must be non-overlapping~\cite{scotti2024clusteringaggregationalgorithmneutralatoms}. In this representation, each valid clustering corresponds to an independent set in the graph, and finding a maximum independent set provides a robust consensus clustering that mitigates the limitations of individual algorithms~\cite{Li2012ClusteringAA}. The application is designed to encourage balanced solutions by penalising both disproportionately large and very small clusters.
The aggregation task is therefore mapped to a MIS problem, to be solved as discussed in Section~\ref{sec:exp_vqpu}.
Expressing part of the problem as a QUBO makes it compatible with a variety of quantum approaches. Since the focus of these experiments is on evaluating scheduling strategies rather than quantum algorithmic performance, we solve the QUBO using Simulated Annealing (SA)~\cite{optimizationBySimulatedAnnealing} on a dedicated node, which serves as a classical proxy for a quantum annealer while preserving the application's hybrid structure and resource usage patterns. Once a solution is obtained, MPI rank 0 computes the silhouette score to evaluate clustering quality by measuring the cohesion of data points within their assigned clusters and their separation from other clusters. Higher silhouette values indicate well-formed and well-separated clusters. Based on this score, the application either reruns the clustering algorithms with updated configurations to improve the result or terminates the process.

\subsection{Test infrastructure}\label{sec:infra}

The two setups are executed across three platforms. For the graph coloring problem, we used Lagrange, a 5-qubit IQM quantum computer, while for the clustering aggregation problem, we used Qluster, a dedicated Slurm testbed and emulation cluster, as well as Leonardo, a production EuroHPC Tier-0 system. Their characteristics are described below.

\subsubsection{Lagrange}
For the vQPU experiment, the quantum workload of each hybrid job was dispatched to the \textit{Lagrange} quantum computer,\footnote{\url{https://docs.quantum.linksfoundation.com}} a gate-based, 5-qubit Spark-model device produced by IQM, hosted at Politecnico di Torino, and managed by Links Foundation. \textit{Lagrange} can be accessed via the standard IQM API, enabling users to submit quantum circuits using several quantum software development kits (SDKs), such as Qiskit, Cirq, and Qrisp.

\subsubsection{Qluster}
We set up a dedicated cluster to test our clustering aggregation problem, using Slurm version 23.02.7. The cluster represents a plausible HPC-QC integration scenario at scale, comprising a frontend node, a master node, and two worker partitions. The frontend node and the master node are virtualised x86\_64 machines with $8 \, \mathrm{GB}$ of memory each. The first partition, named \emph{compute}, consists of three CPU-only nodes. Each compute node contains two AMD EPYC 7543 CPUs and $256 \, \mathrm{GB}$ of DDR4 memory. The second partition, named \emph{quantum}, acts as a quantum emulator. This node contains two AMD EPYC 7282 CPUs, $512 \, \mathrm{GB}$ of DDR4 memory. However, due to the memory requirements of the clustering algorithms, we launch our application requesting one or more classical nodes and one task per node.
The application source code is compiled with GCC version 11.4.1 and the distribution of classical workloads across the compute partition is achieved with Open MPI version 5.0.6. Submitting quantum jobs to the quantum partition is carried out with HyperQueue~\cite{hyperqueue} version 0.21.1. StreamFlow version 0.2.0.dev12 is used to validate the workflow approach.

\subsubsection{Leonardo}
The pre-exascale EuroHPC Tier-0 system Leonardo,\footnote{\url{https://www.hpc.cineca.it/systems/hardware/leonardo}}
 hosted at CINECA, features two types of compute nodes, exposed as separate partitions. The Data-Centric General Purpose (DCGP) partition comprises 1,536 nodes, each equipped with two Intel Sapphire Rapids processors (112 cores total) and 512~GiB of DDR5 memory. The Booster partition includes 3,456 nodes, each featuring an Intel Ice Lake 32-core CPU, 512~GiB of DDR4 memory, and four NVIDIA Ampere GPUs. Both partitions are interconnected through a 200~Gbps NVIDIA Mellanox HDR InfiniBand network, and job scheduling is managed by Slurm~23.11.10.
On Leonardo, DMR has been built and installed through the Spack package manager,\footnote{\url{https://spack.readthedocs.io/en/v0.22.2}} using a custom recipe with \texttt{gcc/12.2.0} as the main compiler, \texttt{cmake/3.27.9} as the build system, and \texttt{openmpi/5.0.8}, \texttt{prrte/4.0.0} and \texttt{pmix/6.0.0} as dependencies. The installation has been exposed as a module through the Environment Modules framework.
Workflow experiments have been orchestrated by StreamFlow version 0.2.0.dev12.

\subsection{Results}\label{sec:results}
We present the experimental results for each strategy separately, beginning with the graph coloring vQPU experiments on real quantum hardware (Section~\ref{sec:results_vqpu}), followed by the clustering and aggregation experiments with malleability and workflows on emulated QPUs (Section~\ref{sec:results_mw}).
\subsubsection{VQPUs}\label{sec:results_vqpu}
\begin{figure}
    \centering
    \includegraphics[width=0.9\linewidth]{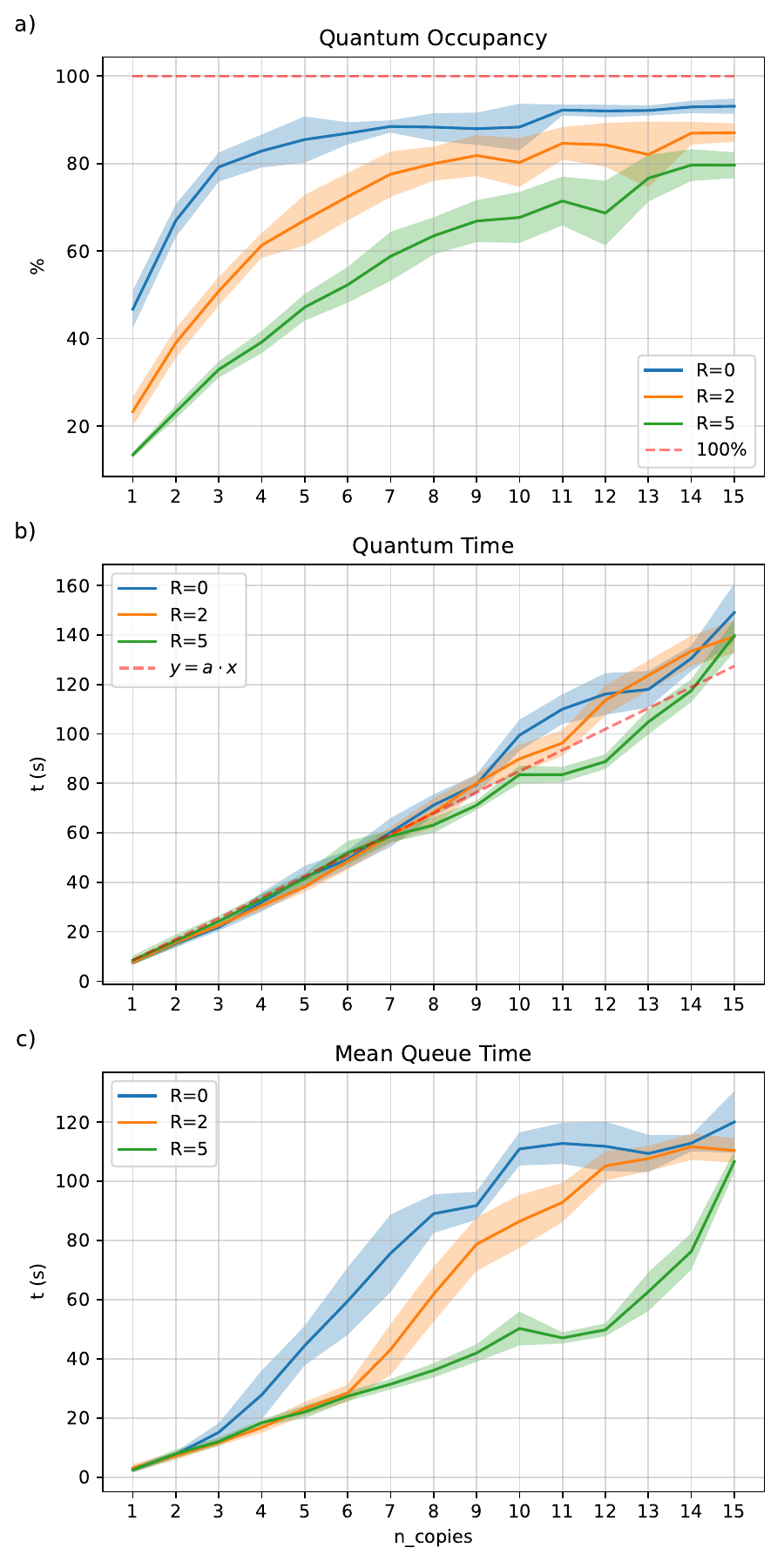}
    \caption{The plots depicting \textit{Quantum Occupancy} (a), \textit{Quantum Time} (b) and \textit{Mean Queue Time} (c) with respect to an increasing number of concurrent GC replicas ($n\_copies$). The labels $R \in \{0, 2, 5\}$ refer to runs with a different quantum-classical workload ratio set, as explained at the beginning of Section \ref{sec:results_vqpu}. The $y=a\cdot x$ dashed line in Sub-figure (b) is used to highlight the linearity relation of the data.}
    \label{fig:vqpu_res_1}
\end{figure}

\begin{figure*}
    \centering
    \includegraphics[width=0.9\linewidth]{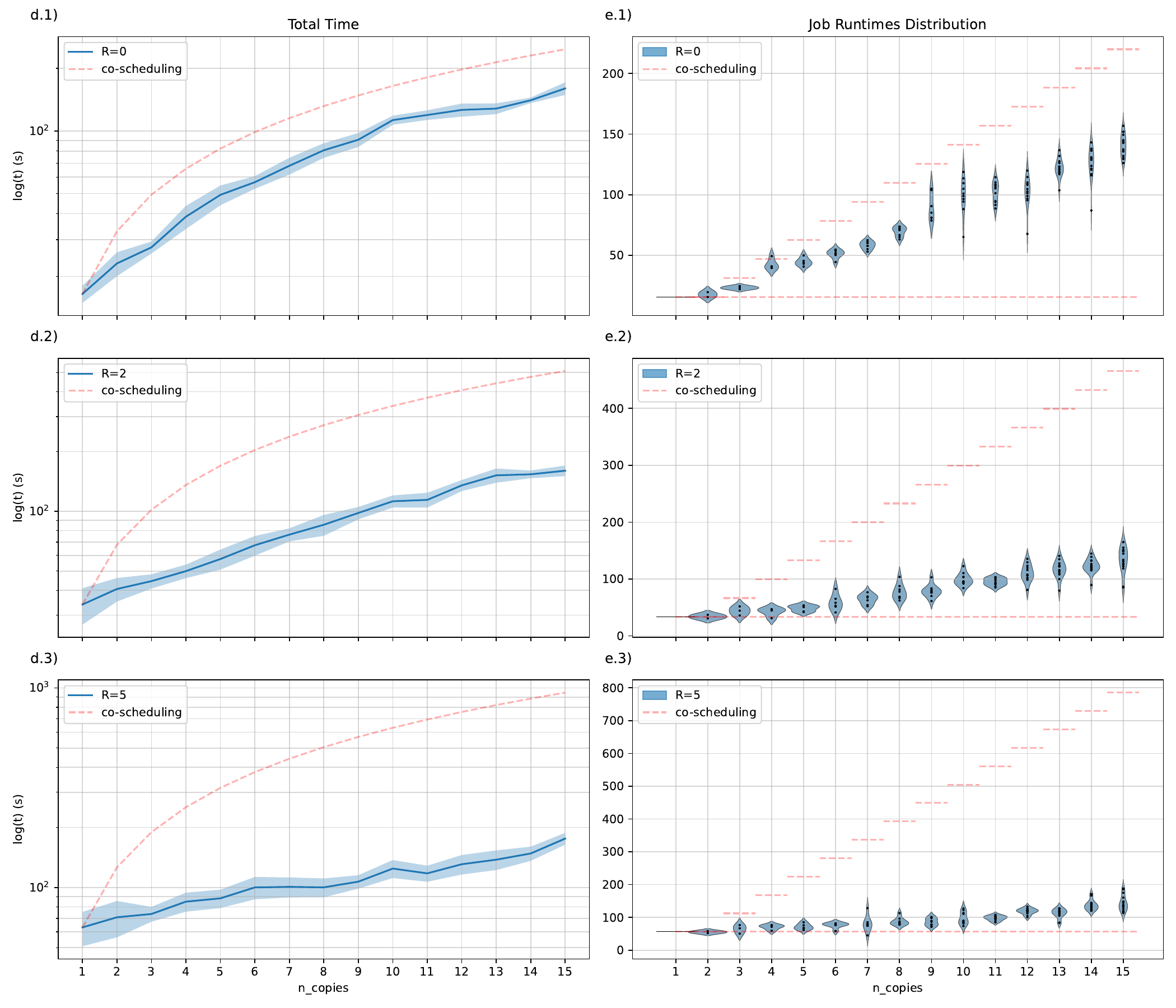}
    \caption{The plots depicting \textit{Total Time} and \textit{Job Runtimes Distribution} metrics. The vertical axis is set on a log scale for plots (d.1-3), while a regular linear scale is used for plots (e.1-3).}
    \label{fig:vqpu_res_2}
\end{figure*}

The experimental results are presented in Figures \ref{fig:vqpu_res_1} and \ref{fig:vqpu_res_2}, where average values (solid lines) and standard deviations (shaded areas) are reported over 10 independent sample runs. It is important to note that, although all jobs are executed using the same input graph, the intrinsic probabilistic nature of the quantum component introduces a degree of variability in the time-to-solution of the problem, as reflected by the observed standard deviation values. Nevertheless, the mean trends exhibit a consistent behaviour with the hypotheses underlying the vQPUs approach (see Figure~\ref{fig:vqpu_scheduling}).

As mentioned in Section~\ref{sec:exp_vqpu}, job runs were executed with an increasing sleep time in order to enlarge the ratio between classical and quantum workloads. This sleep time was introduced at the end of each quantum circuit execution, with its value set proportional to the corresponding quantum execution time. The proportionality factor is denoted by $R$ in the curves shown in the figure. For example, a value of $R=2$ indicates a sleep time twice as long as the quantum execution time, whereas $R=0$ denotes that no sleep mechanism has been set. The variable on the horizontal axis of all plots, i.e., $n\_copies$, denotes the number of GC replicas executed concurrently. This parameter is used to observe how the quantities of interest evolve as the number of hybrid jobs submitted by hypothetical users to the cluster increases.

In Sub-figure \ref{fig:vqpu_res_1} (a), the \textit{quantum occupancy} is reported, defined as the percentage of the total execution time during which the QPU is actively used to run circuits while all jobs are processed on the cluster. As $n\_copies$ increases, the quantum occupancy also rises. This behaviour can be explained by the larger number of concurrent processes submitting circuits and, in agreement with the assumption illustrated in Figure \ref{fig:vqpu_scheduling}, the QPU is able to execute part of the quantum workload of one process while, at the same time, the cluster nodes handle the classical workload of other processes. In this way, both quantum and classical resources are efficiently utilised at cluster level, thereby minimizing waiting times for QPU access overall. The occupancy value eventually reaches a saturation point, beyond which QPU utilisation cannot be further optimized and the submission of additional jobs progressively diminishes the returns obtained by employing the vQPUs strategy with respect to co-scheduling one. This phenomenon will be further examined in the analysis of the following metrics. It is worth noting that the saturation level is reached more gradually as the ratio between classical and quantum workload increases, i.e., for higher values of $R$.

Sub-figures \ref{fig:vqpu_res_2} (d.1-3) report the \textit{total time}, defined as the elapsed time between the start of the first GC replica and the completion of all replicas on the cluster. For comparison with co-scheduling, a dashed curve represents the value of the trivial exclusive scheduling represented in Figure~\ref{fig:coscheduling}
$$
T_{co-scheduling}=T_{exec}(n\_copies=1)\times n\_copies
$$
Here, $T_{exec}(n\_copies=1)$ corresponds to the execution time of a single hybrid job. Consequently, $T_{co-scheduling}$ indicates approximately the time that the cluster would require to process $n\_copies$ GC replicas sequentially, as if quantum resources were allocated exclusively to each of them. As shown by the curves, the overall execution time, plotted on a logarithmic scale, decreases significantly thanks to the use of vQPUs. Moreover, as the workload imbalance increases, the advantage of vQPUs over co-scheduling becomes more pronounced, since the gap between the corresponding curves grows accordingly.

Sub-figure \ref{fig:vqpu_res_1} (b) shows the \textit{quantum time}, i.e., the time actually spent by the quantum computer executing the circuits submitted by all hybrid jobs. This metric does not take into consideration queuing times. Since the \textit{quantum time} is unaffected by the classical workload, the curves corresponding to different values of $R$ overlap. As expected, given the fixed amount of QPUs, the quantum time increases linearly with the number of GC replicas managed by the cluster, and it is not affected by the scheduling policies.

Sub-figure \ref{fig:vqpu_res_1} (c) reports the \textit{mean queue time}, defined as the average cumulative waiting time that each hybrid job undergoes before gaining access to the quantum computer for circuit execution. This metric exclusively refers to the allocation of quantum resources and is independent of the management of classical resources on the cluster via Slurm. As shown by the curves, the \textit{mean queue time} increases with the number of GC replicas, since a larger number of jobs compete for access to the quantum computer. The curves display an upward trend in correspondence with the saturation of the \textit{quantum occupancy}, leading to longer queue times and a consequent reduction in the performance advantage provided by vQPUs. When the maximum value of \textit{quantum occupancy} is eventually reached, the curves reverse their trends. At this point, the addition of a new job increments the cumulative waiting time by a fixed amount, which is balanced by the total number of jobs in the computation of the mean, leading to a slowly increasing value. Similar to the trends observed for the \textit{total time}, the benefit of using vQPUs becomes more pronounced as the workload imbalance increases, i.e., the average waiting time experienced by each user decreases for larger values of $R$.

In Sub-figures~\ref{fig:vqpu_res_2} (e.1-3), the \textit{jobs' runtime distributions} are reported. Each violin plot represents the distribution of execution times across the set of concurrent jobs, with the individual runtimes indicated as dots within each plot. For comparison with co-scheduling, the corresponding \textit{total times} are also shown as the upper dashed lines, while the lower one represents the theoretical duration of the first GC job submitted. As highlighted in Section~\ref{sec:methods_vqpu}, the vQPU approach aims to optimise resource utilisation at the cluster level, which may, however, result in longer runtimes for some jobs, since quantum resources are shared among all users. Given that the experiment simulates the worst-case scenario in which all jobs are launched simultaneously, the results reveal a subset of jobs, visible in the bottom section of each violin plot, for which the delay relative to their co-scheduling counterparts is more pronounced. As a reference, the distance between the lower dashed line and lowest dot of each plot can be considered. Nevertheless, when considering the overall distribution, the adoption of vQPUs leads to a significant acceleration for the majority of concurrent jobs, yielding a reduced \textit{total time}, in line with the results in Sub-figures~\ref{fig:vqpu_res_2} (d.1-3).

\noindent\textbf{Take-home message.} Time-multiplexing a physical QPU through vQPUs transparently increases quantum occupancy and significantly reduces cluster-level execution time, with benefits growing as the classical-to-quantum workload imbalance increases. The approach is most effective for superconducting QPUs with short quantum bursts, and requires zero modifications to user applications, though individual jobs may experience slightly longer runtimes due to shared QPU access.

\subsubsection{Dynamic resource management and workflows}\label{sec:results_mw}
\paragraph{Common metrics.}
We execute a set of comparative experiments to evaluate the effectiveness of the \emph{malleability} and \emph{workflow} approaches described in Sections~\ref{sec:methods_malleability} and~\ref{sec:methods_wf}, respectively, for hybrid HPC-QC scheduling on both Qluster and Leonardo. We begin by analysing a configuration that adopts a traditional resource management strategy, allocating classical resources for the entire duration of the hybrid job and offloading the quantum subroutine to the emulated QPU. This configuration is referred to as the \emph{baseline} throughout the rest of the Section, as it resembles the current offloading scheme adopted with cloud-based quantum machines. We then compare the baseline results to those obtained with the malleability and workflows approaches.
On Qluster, we use HyperQueue as an abstraction layer to schedule emulated quantum jobs for the baseline and malleability configurations, whereas the workflow approach independently allocates a node from the quantum partition only when required. On Leonardo, we adopt a slightly revised scheduling approach to allow concurrent usage of any node from the quantum emulation partition, accomplished by removing the HyperQueue metascheduler and letting jobs directly submit other Slurm jobs.

For all three scheduling approaches, we evaluate and compare the following metrics:
\begin{itemize}
    \item \textit{Wall time}: the time employed to execute the whole simulation, considering also the time spent by Slurm to initialise and finalise the job(s);
    \item \textit{Classical resource usage}: the sum of the product between the number of classical nodes used in an interval and the duration of the interval (in seconds). Note that we disregard here the quantum resources, as their usage is constant across the three scenarios.
\end{itemize}

The dataset used in the clustering application consists of 80,000 2D points generated via \texttt{make\_blobs} from scikit-learn.\footnote{\url{https://scikit-learn.org/stable/modules/generated/sklearn.datasets.make_blobs.html}} The application code is written to permit reproducible results. In particular, given the described dataset, each workload run ends after the fourth loop iteration, as the combined clusterings achieve a silhouette score over 0.8.
We profile two versions of the application to better understand how different quantum technologies might suit specific scheduling approaches. A first version executes quantum jobs in a fraction of a second, thus akin to superconducting QPUs. A second version adds an artificial delay of two minutes during the quantum job, representative of the longer execution times typical of neutral-atom QPUs.

\paragraph{Qluster experiments}
We start from the executions with no resource contention, i.e., with no other jobs in the cluster queue, and with the two-minute-long quantum jobs, reproducing the behaviour of a neutral-atom machine. We average the metrics from five runs for each strategy.
Table~\ref{table:medium_jobs_combined} shows the results of the first experiment. The baseline approach is the fastest, but it is the least efficient in terms of resource usage. The workflow approach performs poorly in terms of wall time, since
the WMS requests new SLURM resources at every step, and it can wait up to $1 \, \mathrm{s}$ to detect a job's termination due to its polling logic, resulting in approximately $2 \, \mathrm{s}$ of overhead per step; conversely, it achieves the lowest resource usage with minimal node-second consumption. The malleability approach acts as a compromise between the other two.
Its overhead depends on the requested reconfiguration: expansions require new allocations via SLURM and take approximately $1.5 \, \mathrm{s}$, whereas shrinks reuse existing allocations and take only around $0.7 \, \mathrm{s}$ to restore the application context.
In the absence of resource contention, both malleability and workflow approaches primarily conserve valuable computational resources with a negligible impact on time-to-solution.

\begin{figure*}[t]
    \centering
    \includegraphics[width=0.62\linewidth]{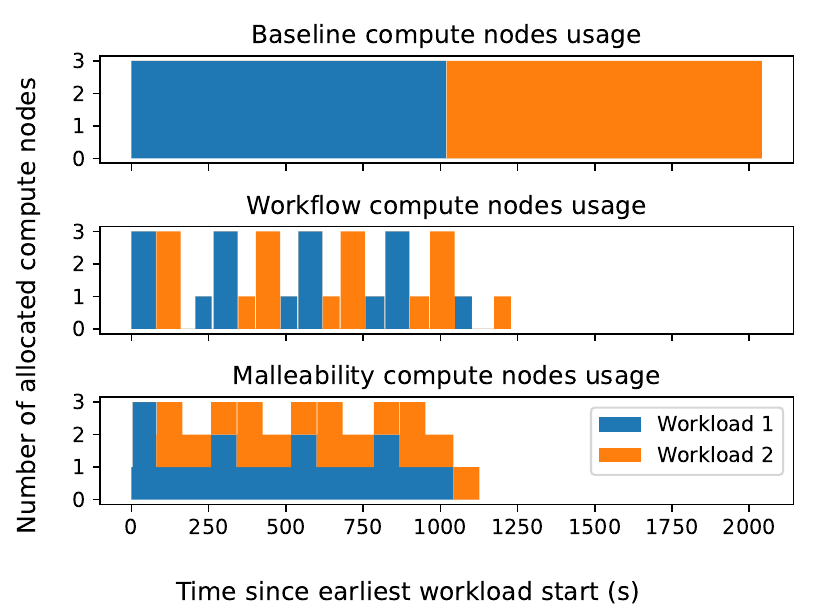}
    \caption{Timeline of cumulative usage of classical nodes on our Slurm compute partition when launching two concurrent workloads (in blue and orange, respectively) with two minutes long quantum jobs.
    Each subplot refers to a different scheduling approach, i.e. baseline, workflow, and malleability.}
    \label{fig:comparison_dual_workloads_2min_sleeps}
\end{figure*}

For our second experiment, we run two concurrent workloads using all the approaches, again under a queue empty from other submissions and by emulating two-minutes-long quantum jobs.
Table~\ref{table:medium_jobs_combined} contains the results of this second experiment, averaged over five runs each. Note that the wall time here refers to the elapsed time between the beginning of the first starting workload and the completion of the latest ending workload, thus considering two complete end-to-end simulations. The baseline approach is, in this case, the worst-performing one. The other approaches can interleave their execution, finishing earlier and using fewer resources, as shown in Figure~\ref{fig:comparison_dual_workloads_2min_sleeps}.
The difference between malleability and workflow results resides in the need for the former to have at least one MPI process to remain active at all times, even during the quantum phase when computations are offloaded to the QPU. While this may seem like an overhead, it offers a clear advantage in our case: when the code returns from the quantum phase, execution can resume immediately, allowing the simulation to proceed even if not all originally requested resources are available.

We then execute the experiments above without artificial delays in the quantum phase. The results of these new experiments can be seen in Table~\ref{table:single_workload_no_sleep}. For baseline and workflow, the wall time and resource usage of the dual concurrent execution almost double the values from the single execution. This means the classical resources are highly contested, with little space to optimise scheduling. The malleability solution, on the other hand, completes both simulations in a time comparable to that of a single execution ($648.61 \, \mathrm{s}$ vs $549.60 \, \mathrm{s}$), demonstrating its ability to manage and adapt to varying resource demands dynamically. However, this result comes from the intrinsic properties of the algorithms executed during the classical phase (with k-means taking significantly less time than the other two), so they do not represent a trend. In general, scenarios involving short quantum jobs and no resource contention should show limited resource savings when using either workflow or malleability approaches.

Lastly, we observe low variance for each Qluster experiment regarding both wall time and resource usage. This is expected, since the target cluster is a controlled environment with no external users interfering with our workloads.

\begin{table}[]
\centering
\caption{Qluster: executions with two-minute-long quantum jobs.}
\label{table:medium_jobs_combined}
\begin{tabular}{lccc}
\toprule
\begin{tabular}[c]{@{}c@{}}Execution\\Type\end{tabular} & Mode & \begin{tabular}[c]{@{}c@{}}Wall time\\ {[}seconds{]}\end{tabular} & \begin{tabular}[c]{@{}c@{}}Resource usage\\ {[}node-seconds{]}\end{tabular} \\
\midrule
Baseline       & Single & 1019.58 $\pm$ 0.85 & 3058.74 $\pm$ 2.56 \\
Workflow       & Single & 1057.80 $\pm$ 6.02 & 1161.20 $\pm$ 6.94 \\
Malleability   & Single & 1029.06 $\pm$ 1.54 & 1647.75 $\pm$ 1.54 \\
\midrule
Baseline       & Double & 2038.43 $\pm$ 0.96 & 6115.30 $\pm$ 2.89 \\
Workflow       & Double & 1226.00 $\pm$ 1.58 & 2324.00 $\pm$ 3.39 \\
Malleability   & Double & 1127.65 $\pm$ 1.18 & 2817.73 $\pm$ 1.27 \\
\bottomrule
\end{tabular}
\end{table}

\begin{table}[]
\centering
\caption{Qluster: executions with short ($<1$ second) quantum jobs.}
\label{table:single_workload_no_sleep}
\begin{tabular}{lccc}
\toprule
\begin{tabular}[c]{@{}c@{}}Execution\\Type\end{tabular} & Mode & \begin{tabular}[c]{@{}c@{}}Wall time\\ {[}seconds{]}\end{tabular} & \begin{tabular}[c]{@{}c@{}}Resource usage\\ {[}node-seconds{]}\end{tabular} \\
\midrule
Baseline       & Single & 539.44 $\pm$ 0.53 & 1618.33 $\pm$ 1.60 \\
Workflow       & Single & 569.00 $\pm$ 3.94 & 1148.00 $\pm$ 1.87 \\
Malleability   & Single & 549.60 $\pm$ 1.86 & 1168.29 $\pm$ 1.81 \\
\midrule
Baseline       & Double & 1076.98 $\pm$ 1.79 & 3230.95 $\pm$ 5.37 \\
Workflow       & Double & 1089.00 $\pm$ 1.00 & 2324.00 $\pm$ 4.24 \\
Malleability   & Double & 648.61 $\pm$ 2.08 & 1622.63 $\pm$ 1.05 \\
\bottomrule
\end{tabular}
\end{table}

\FloatBarrier
\paragraph{Leonardo experiments}

Having previously demonstrated how our solution performs in a controlled environment (Qluster), we now take advantage of access to the production system Leonardo to replicate the same tests under real-world conditions. In order to cope with the volatility of large-scale productions systems, we decided to launch a larger number of runs (32). This turned out to be necessary since, due to a cluster-wide Slurm issue, acknowledged by the system administrators and unrelated with our set-up, a number of runs failed. The times we report are averaged over 29 runs in the short quantum‑job experiment, and 10 in the two‑minute quantum‑job experiment.
Numerical results are reported in Table~\ref{table:baseline_vs_dmr_vs_wf_medium} and Table~\ref{table:baseline_vs_dmr_vs_wf_short}. Consistent with the Qluster results, baseline executions on Leonardo are the fastest but the most wasteful in terms of resource usage. However, we observe substantially higher variances for both wall time and resource usage, which is expected given that Leonardo is a multi-tenant production cluster where workloads from other users affect scheduling and resource availability. In particular, the short quantum job experiments exhibit high standard deviations, caused by a high cluster load during some of the runs.
Workflow emerges as a winner both in terms of resource consumption (up to 64\% less resources than baseline) and resource consumption variance in this kind of scenario. The workflow manager can wait for new nodes to be allocated without consuming resources, while malleability and baseline constantly keep at least one node active and suffer from delays in the quantum emulation queue. Average resource usage is thus lower than baseline.
The workflow manager does not consume cluster resource per se, since it can run on any node able to reach the target Slurm cluster. In our case, we launched StreamFlow on the \texttt{lrd\_all\_serial} partition, which does not consume budget resources on Leonardo. This also means that workflow experiments typically started some minutes after the other approaches, since they needed one extra core from the \texttt{lrd\_all\_serial} partition. Unfortunately, the cluster queue status can change in a matter of minutes. We measured at least two cases of significant differences between approaches due to the starting delay and the ever varying state of the queue.
Nonetheless, the wall times of all approaches are approximately within their standard deviation, on average. The real benefits of our proposed approaches are the improvements in resource usage, which is the metric that we are trying to optimise.
DMR also demonstrates a clear advantage in resource consumption when compared to the baseline (up to $-45.7\%$ node-seconds), as it releases nodes while waiting for the completion of quantum jobs and during the serial silhouette score computation. The wall time of malleable jobs is slightly larger (up to 5\%), since each reconfiguration introduces a small overhead. On the whole, DMR achieves satisfactory results in reducing resource consumption, without significantly penalising execution times.

Both the short quantum jobs and the two-minutes-long quantum jobs experiments lead to the same conclusions. However, both approaches become more efficient as quantum job times increase. For example, taking the two extremes, we can observe how workflows are $1.94\times$ more efficient than the baseline with short quantum jobs and $2.78\times$ more efficient than the baseline with two-minute quantum jobs. Thus, optimised scheduling strategies such as malleability and workflows become even more effective with hybrid applications where the quantum part takes a large portion of the whole execution.

\noindent\textbf{Take-home message.} Both malleability and workflow decomposition substantially reduce classical resource consumption (up to 45.7\% and 64\% respectively) compared to the static baseline, with workflow achieving the lowest resource footprint and variance, and malleability offering a good balance between wall time and efficiency. Their benefits become increasingly pronounced as the quantum phase duration grows, making them particularly well suited for hybrid workloads targeting neutral-atom or other longer-execution-time quantum technologies.

\begin{table}[]
\centering
\caption{Leonardo: executions with two-minute-long quantum jobs.}
\label{table:baseline_vs_dmr_vs_wf_medium}
\begin{tabular}{lll}
\toprule
\begin{tabular}[c]{@{}l@{}}Execution\\ type\end{tabular} & \begin{tabular}[c]{@{}l@{}}Wall time\\ {[}seconds{]}\end{tabular} & \begin{tabular}[c]{@{}l@{}}Resource usage\\ {[}node-seconds{]}\end{tabular} \\
\midrule
Baseline & 1126.10 $\pm$ 21.20                                            & 3378.30 $\pm$ 63.59                                                     \\
\midrule
DMR      & \begin{tabular}[c]{@{}l@{}}1151.50 $\pm$ 40.63\end{tabular} & 1835.60 $\pm$ 41.74                                                     \\
\midrule
Workflow & 1127.90 $\pm$ 20.93                                            & 1216.80 $\pm$ 6.55                                                      \\
\bottomrule
\end{tabular}
\end{table}

\begin{table}[]
\centering
\caption{Leonardo: executions with short ($<1$ second) quantum jobs.}
\label{table:baseline_vs_dmr_vs_wf_short}
\begin{tabular}{lll}
\toprule
\begin{tabular}[c]{@{}l@{}}Execution\\ type\end{tabular} & \begin{tabular}[c]{@{}l@{}}Wall time\\ {[}seconds{]}\end{tabular} & \begin{tabular}[c]{@{}l@{}}Resource usage\\ {[}node-seconds{]}\end{tabular} \\
\midrule
Baseline                                                 & 803.69 $\pm$ 516.77                                               & 2411.07 $\pm$ 1550.30                                                       \\
\midrule
DMR                                                      & 844.03 $\pm$ 529.89                                               & 1517.28 $\pm$ 538.12                                                        \\
\midrule
Workflow                                                 & 818.48 $\pm$ 420.97                                               & 1241.34 $\pm$ 22.79                                                         \\
\bottomrule
\end{tabular}
\end{table}

\section{Conclusion \& Perspectives}\label{sec:summary}
This work presented and experimentally evaluated three complementary scheduling strategies for hybrid HPC-QC environments: virtual QPUs (vQPUs), dynamic resource management through MPI malleability, and workflow decomposition. Each strategy targets a different level of the software stack and addresses a different facet of the resource scheduling challenge, as illustrated in Figure~\ref{fig:overview}. In the following, we summarise the main findings for each approach and discuss their complementarity.

The vQPU approach operates at the infrastructure level, enabling time-based multiplexing of a physical QPU among concurrent hybrid jobs without requiring any modification to user applications. Our experimental campaign on a production HPC cluster with a real quantum computer confirmed that this strategy effectively increases QPU utilisation, leading to significant reductions in overall execution time at the cluster level. The performance gains become increasingly pronounced as the imbalance between classical and quantum workloads grows, making the approach particularly well suited for scenarios involving superconducting quantum hardware, where quantum execution bursts are short relative to classical processing phases. Despite the expected overhead due to contention for QPU access, the observed trends indicate that scaling the number of vQPUs does not significantly increase the average job queue time beyond a certain saturation threshold, suggesting that the strategy remains robust under varying hybrid job submission loads.

The malleability and workflow approaches both target the scheduling of resources from the programming model perspective, but they differ in their integration requirements and operational trade-offs. Malleability, implemented through the DMR framework, operates at the scheduler and runtime level: it allows an existing MPI application to dynamically release and reclaim classical nodes, and possibly quantum, during execution, requiring only a limited number of additional code lines without refactoring the overall application structure. On Qluster, malleability achieved a good balance between wall time and resource consumption, and it proved especially effective under resource contention with concurrent workloads. On Leonardo, DMR reduced resource consumption by up to 45.7\% compared to the baseline, at the cost of a modest increase in wall time (up to 5\%) due to reconfiguration overheads.

The workflow approach, implemented through StreamFlow, operates at the orchestration level: it structures the application as a directed graph of tasks, delegating resource allocation for each step to the WMS. This design allows computational resources to be provisioned only when a task is actively running, which yields the lowest resource consumption among the three approaches (up to 64\% fewer node-seconds than the baseline on Leonardo) and the lowest variance in resource usage. On the other hand, the overhead of repeated resource requests through Slurm increases wall time in the absence of contention, and the WMS introduces a startup delay that complicates direct wall time comparisons in a production environment. Unlike malleability, newly designed applications can be directly developed in a bottom-up fashion as a workflow, and workflow managers such as StreamFlow are general-purpose tools that can target multiple environments (HPC, cloud, local) and programming languages, offering greater versatility. Table~\ref{tab:comparison} summarises the key characteristics of the three strategies across the dimensions most relevant to deployment decisions.

\begin{table*}[t]
\centering
\caption{Comparative summary of the three scheduling strategies.}
\label{tab:comparison}
\small
\begin{tabular}{@{}lccc@{}}
\toprule
 & \textbf{vQPUs} & \textbf{Malleability (DMR)} & \textbf{Workflow (StreamFlow)} \\
\midrule
\textbf{Software stack level} & Infrastructure & Scheduler / runtime & Orchestration \\
\midrule
\textbf{Overhead on individual job}
  & Quantum queue time & Reconfiguration cost & Slurm queue \\
  & (small and tunable) & (small for shrinking) & (variable) \\
\midrule
\textbf{Resource efficiency} & & & \\
\quad -- single job & Neutral & Moderate savings & High savings \\
\quad -- cluster-wide & High (QPU utilisation) & Moderate--high savings & High savings \\
\midrule
\textbf{Best workload profile}
  & Short QC, & Medium-long QC, & Medium-long QC, \\
  & medium-long HPC & medium-long HPC & medium-long HPC \\
\midrule
\textbf{Applicability and impact}
  & Any hybrid job & MPI applications & WMS-ready applications \\
  & (transparent) & (few LOC changes) & (may require workflow decomposition) \\
\bottomrule
\end{tabular}
\end{table*}

From a practical standpoint, the choice among the three strategies depends on the specific workload characteristics and operational constraints. For HPC centres deploying superconducting QPUs with short quantum execution bursts, vQPUs offer the highest impact with zero application modifications. For existing MPI applications that need to coexist with quantum offloading on shared clusters, malleability through DMR provides a low-overhead path to improved resource efficiency. For newly developed applications or heterogeneous workflows spanning multiple execution environments, the workflow approach offers the greatest flexibility and the lowest resource footprint, at the cost of higher integration effort and sensitivity to scheduler latency. Importantly, since our rationale focuses primarily on the relative weight of the classical and quantum components of the workload rather than the specific quantum technology in use, these findings are expected to remain robust as quantum hardware evolves and technical specifications change.

The three strategies are not mutually exclusive. As discussed in Section~\ref{sec:methodology}, they occupy distinct but partially overlapping regions in the space defined by time-scale granularity and programming model transparency (Figure~\ref{fig:overview}). In practice, a user may leverage workflow decomposition or malleability without being aware that the targeted QPU is also being shared through vQPUs. Similarly, malleability and workflow approaches can be combined: a workflow manager might oversee the global execution of the application while selected malleable components are automatically scaled up or down depending on the cluster state.

To the best of the authors' knowledge, this work also represents the first practical demonstrations of HPC-QC hybrid jobs fully executed on hardware deployed in Italy, contributing to bridging the gap between theoretical approaches and real-world deployments.

\subsection{Future work}\label{sec:future}
Several directions for future work emerge from our findings. The vQPU approach depends heavily on the specific HPC centre set-up, policies, and machine API; this work focused on evaluating the scheduling performance in a realistic usage scenario, while the details of the Slurm configuration and the additional services necessary to support this solution in a production environment (e.g., authorisation, billing, quotas) are beyond the scope of this paper and will be addressed in future work.

Another key aspect will be to investigate how the performance of the proposed approaches depends on specific algorithmic features that alter the interplay between quantum and classical resources, such as variational versus non-variational algorithms, thereby expanding our evaluation to a more diverse set of workloads.

For malleability, we aim to design and experiment with customised reconfiguration policies to enable the orchestration of more complex workloads. Concurrently, the integration of newer DMR versions that dynamically detect node readiness may further reduce reconfiguration overheads.
Additionally, exploring the interplay between malleability and communication efficiency in HPC-QC workloads represents a promising research direction. For workflow decomposition, extending the experimental evaluation to scenarios involving actual quantum hardware (rather than emulated QPUs) and investigating the combination of workflow orchestration with malleable sub-tasks are natural next steps. More broadly, evaluating the three strategies under larger-scale, more diverse workloads and with different quantum computing technologies (e.g., neutral atoms, trapped ions) will be essential to further generalise the validation of the presented scheduling methodology.


\section*{Funding}
Funded by the European Union – NextGenerationEU: ICSC National Centre, CN00000013, MUR Act n. 1031 - 17/06/2022. This publication is part of the project PNRR-NGEU which has received funding from the MUR – DM 630/2024.

The authors acknowledge access to the IQM quantum computer Lagrange, jointly acquired by Fondazione LINKS, Politecnico di Torino and Istituto Nazionale di Ricerca Metrologica (INRiM), which was used to perform the hybrid HPC-QC experiments reported in this work. This work has also been supported by CINECA national HPC resources allocation grant IscraC.

BSC researchers have been financially supported by the Ministry of Economic Affairs and Digital Transformation of the Spanish Government through the QUANTUM ENIA project call - Quantum Spain project, and by the Ministerio para la Transformación Digital y de la función pública, within the framework of the Plan de Recuperación Transformación y Resiliencia, and by the European Union --- NextGenerationEU and the framework of the Digital Spain 2026 Agenda.
Antonio J. Peña was partially supported by the Ramón y Cajal fellowship RYC2020-030054-I funded by MCIN\slash AEI\slash 10.13039\slash 501100011033 and by ``ESF Investing in your future''.
The views and opinions expressed are solely those of the authors and do not necessarily reflect those of the European Union. Neither the European Union nor the European Commission can be held responsible for them.

\section*{Declaration of generative AI and AI-assisted technologies in the manuscript preparation process}
During the preparation of this work the author(s) used Claude for grammar review and language polishing. After using this tool/service, the author(s) reviewed and edited the content as needed and take(s) full responsibility for the content of the published article.

\section*{Data Availability}
The code used for the Qluster experiments is available on GitHub~\cite{clustering-mis}. We applied minor adjustments to run the clustering aggregation experiments on Leonardo to target the proper SLURM partitions and to load appropriate modulefiles. The code used for the vQPU experiment is available publicly~\cite{vqpu_gitlab}.

\bibliographystyle{cas-model2-names}

\bibliography{bibliography_revised}

\end{document}